\documentclass[aps,prb,twocolumn,showpacs, superscriptaddress, floatfix]{revtex4}
\usepackage{graphicx}

\begin{document}
\title{Ultrafast carrier dynamics and radiative recombination in multiferroic BiFeO$_{3}$}

\author{Y. M. Sheu}
\affiliation{Center for Integrated Nanotechnologies, MS K771, Los Alamos National Laboratory, Los Alamos, New Mexico 87545, USA}
\author{S. A. Trugman}
\affiliation{Center for Integrated Nanotechnologies, MS K771, Los Alamos National Laboratory, Los Alamos, New Mexico 87545, USA}
\author{Y.-S. Park}
\affiliation{Center for Integrated Nanotechnologies, MS K771, Los Alamos National Laboratory, Los Alamos, New Mexico 87545, USA}
\author{S. Lee}
\affiliation{Rutgers Center for Emergent Materials and Department of Physics and Astronomy, Rutgers University, 136 Frelinghuysen Rd.,
Piscataway, New Jersey 08854, USA}
\author{H. T. Yi}
\affiliation{Rutgers Center for Emergent Materials and Department of Physics and Astronomy, Rutgers University, 136 Frelinghuysen Rd.,
Piscataway, New Jersey 08854, USA}
\author{S.-W. Cheong}
\affiliation{Rutgers Center for Emergent Materials and Department of Physics and Astronomy, Rutgers University, 136 Frelinghuysen Rd.,
Piscataway, New Jersey 08854, USA}
\author{Q. X. Jia}
\affiliation{Center for Integrated Nanotechnologies, MS K771, Los Alamos National Laboratory, Los Alamos, New Mexico 87545, USA}
\author{A. J. Taylor}
\affiliation{Center for Integrated Nanotechnologies, MS K771, Los Alamos National Laboratory, Los Alamos, New Mexico 87545, USA}
\author{R. P. Prasankumar}
\affiliation{Center for Integrated Nanotechnologies, MS K771, Los Alamos National Laboratory, Los Alamos, New Mexico 87545, USA}

\date{\today}

\begin{abstract}
We report a comprehensive study of ultrafast carrier dynamics in single crystals of multiferroic BiFeO$_{3}$.  Using femtosecond optical pump-probe spectroscopy, we find that the photoexcited electrons relax to the conduction band minimum through electron-phonon coupling with a $\sim$1 picosecond time constant that does not significantly change across the antiferromagnetic transition. Photoexcited electrons subsequently leave the conduction band and primarily decay via radiative recombination, which is supported by photoluminescence measurements. We find that despite the coexisting ferroelectric and antiferromagnetic orders in BiFeO$_{3}$, the intrinsic nature of this charge-transfer insulator results in carrier relaxation similar to that observed in bulk semiconductors.

\end{abstract}
 \maketitle

Bismuth ferrite (BFO) is one of the most actively studied multiferroic materials due to its room temperature coexistence of ferroelectric (FE) ($T_{c}$$\sim$1100 K) and antiferromagnetic (AFM) ($T_{N}$$\sim$640 K) orders. Much research has focused on enhancing their weak mutual coupling, particularly by using growth techniques that vary the structure or strain in BFO films.\cite{Wang2003Science, Li2004APL, Dupe2010PRB, Christen2011PRB, Ratcliff2011AM, Prosandeev2011PRB} This could allow both control of magnetism with electric fields and control of electric polarization with magnetic fields, which would lead to a variety of potential applications in optoelectronics, spintronics, and data storage.

Despite the intense research on this material, relatively few studies of its optical properties have been done to date. However, these studies have uncovered several unique phenomena that are linked to FE order in BFO. For example, calculated and measured absorption spectra reveal the optical band gap to be $\sim$2.6-2.8 eV at 300 K,\cite{Xu2009PRB,Pisarev2009RPB,Clark2007APL} arising from the dipole-allowed O 2$p$ to Fe 3$d$ charge transfer (CT) transition. These measurements have also revealed strong absorption edge "smearing"\cite{Ramirez2009PRB, Pisarev2009RPB, Xu2009PRB} which was attributed to low-lying electronic \cite{Pisarev2009RPB,Basu2008APL} or defect states,\cite{Basu2008APL,Hauser2008APL} both of which can strongly impact the ferroelectric response. Terahertz (THz) emission spectroscopy indicates that ultrafast depolarization of the FE order causes the observed emission,\cite{Takahashi2006PRL,Rana2009AM} although the detailed mechanism is not entirely clear. A substantial zero-bias photovoltaic effect has also been observed in BFO with near/above band gap illumination, and the photocurrent preferentially moves along the direction of FE polarization.\cite{Choi2009Science}

Deeper insight into these and other phenomena, as well as their potential for applications (e.g., in determining switching speeds in BFO-based devices), can be gained by tracking carrier dynamics in BFO on an ultrafast timescale,\cite{Chen2011JSNM} particularly after excitation of the $p$-$d$ CT transitions that dominate the near band gap optical response and have been linked to FE properties.\cite{Pisarev2009RPB} This can be done using ultrafast optical spectroscopy (UOS), a technique that is capable of tracking the interplay between carrier, spin and lattice degrees of freedom by using femtosecond optical pulses to photoexcite materials and probing the response in the time domain.\cite{Averitt2002JPCM,Ren2008PRB,Rohit2011Book} In this letter, we use UOS, in conjunction with time-integrated photoluminescence (PL) spectroscopy, to unravel carrier dynamics in BFO after CT excitation.

A BFO (001)$_{pc}$ single crystal with 220 $\mu$m thickness was grown by Bi$_{2}$O$_{3}$ flux. THz time-domain spectroscopic measurements confirmed the existence of a single FE domain.\cite{Talbayev2008APL,Talbayev2011PRB} The optical pump-probe experiment is based on an amplified Ti:sapphire laser producing pulses at a 250 kHz repetition rate with $\sim$100 fs duration and energy $\sim$4 $\mu$J/pulse at a center wavelength of 785 nm (1.58 eV). These pulses are then frequency doubled through second harmonic generation in a BBO crystal, producing UV pump and probe pulses with center wavelengths tunable from 383-398 nm (3.12-3.23 eV) that are focused onto the BFO crystal in a reflection geometry with a variable time delay $t$ between them. This allows us to photoexcite the O 2$p$ to Fe 3$d$ CT transition and probe the resulting photoinduced transient change in reflectivity, $\Delta$R/R ($t$). Photoluminescence signals are generated after photoexcitation by laser pulses with center wavelength 405 nm (3.06 eV) and 50 ps duration at a 10 MHz repetition rate. A fast avalanche photodiode detector is used to collect the PL signal.

\begin{figure}[tb]
\begin{center}
\includegraphics[width=3in]{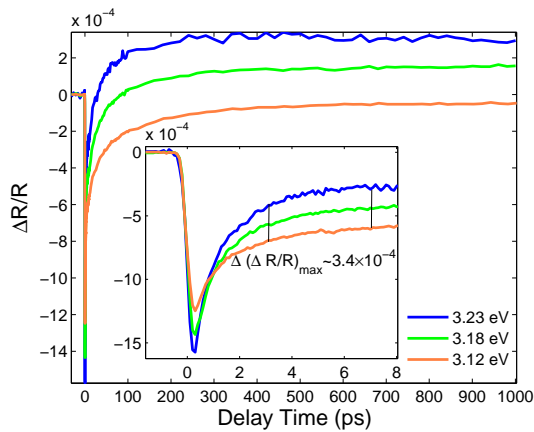}
\caption{ \label{f:SC_Wave} (color online)  Degenerate pump-probe-reflectivity measurements at 3 different photon energies at room temperature. The inset displays the early time dynamics, with two identical vertical bars that represent the maximum difference between the signals at the highest and lowest pump photon energies. The signals reach their maximum difference at $\sim$3 ps.}
\end{center}
\end{figure}

Fig. \ref{f:SC_Wave} shows room temperature pump-probe reflectivity data at three different photon energies. Our data is best fitted with three decay time constants with magnitudes of $\sim$1 ps (fast time constant), $\sim$10-50 ps (intermediate time constant), and $\sim$1-3 ns (slow time constant) at all measured temperatures.  Initially, the pump pulse will excite electron-hole pairs above the band gap, modifying FE order as previously shown.\cite{Takahashi2006PRL,Rana2009AM} After photoexcitation, the non-equilibrium electron-hole pairs bleach the absorption through state filling, resulting in the initial negative $\Delta$R/R ($t$=0) signal. The photoexcited electrons rapidly thermalize amongst themselves and subsequently relax to the conduction band minimum via electron-phonon coupling \cite{carrierNotes} (explained in more detail below).

Subsequent relaxation across the band gap can proceed either through non-radiative or radiative recombination. If carrier relaxation across the gap is primarily due to non-radiative recombination, then all the absorbed energy will eventually be transferred to the lattice as heating; one would thus expect to see a large positive offset in the $\Delta$R/R signal at long time delays, because the time-integrated reflectivity in BFO (at energies near and above the gap) increases with increasing temperature.\cite{Palai2008PRB} In contrast, if carrier relaxation across the gap is dominated by radiative recombination, only the excess energy relative to the band minimum ($\sim$0.4 eV) will contribute to lattice heating, and one would thus expect to see a smaller positive offset in the $\Delta$R/R signal at later times. Our data agrees with the latter case, since if all photoexcited carriers non-radiatively recombined the calculated lattice temperature increase would be $\Delta T_l\sim$4.5 K\cite{dRnotes} for our absorbed pump fluence of $\sim$36 $\mu$J/cm$^{2}$. This corresponds to $\Delta$R/R$\sim$ 5.7$\times$10$^{-3}$, which is much larger than our measured values at long time delays (Fig. \ref{f:SC_Wave}).  Furthermore, by measuring $\Delta$R/R at three different photon energies, one can see that the $\Delta$R/R signals all reach an essentially constant value at long times. If the majority of photoexcited carriers radiatively recombine across the gap, then the 0.11 eV difference between the highest and lowest photon energies compared to the $\sim$2.65 eV gap energy should not be large enough to substantially change the reflectivity signal.  We calculate that the difference in the photoinduced transient reflectivity change for the highest and lowest photon energies in this scenario is $\Delta$($\Delta$R/R)$\sim$2$\times$10$^{-4}$, which agrees well with the $\Delta$($\Delta$R/R)$\sim$3.4$\times$10$^{-4}$ observed in our data (inset of Fig. \ref{f:SC_Wave}).\cite{dRnotes} In addition, we find that the maximum difference in $\Delta$R/R between the three measured photon energies is reached within 3 ps, as shown in the inset of Fig. \ref{f:SC_Wave}. These considerations further support the picture that photoexcited electrons relax to the bottom of the conduction band and transfer their energy to the lattice within 3 ps in BFO. The fast time constant in our data is therefore due to carrier relaxation via electron-phonon coupling.

\begin{figure}[tb]
\begin{center}
\includegraphics[width=2.4in]{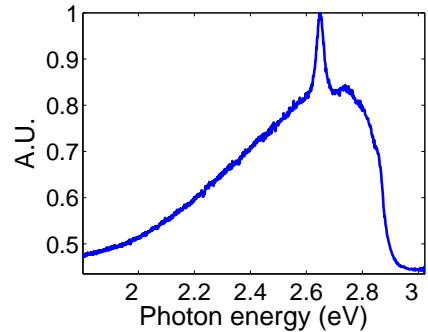}
\caption{ \label{f:PL}
Photoluminescence spectrum measured by 3.06 eV excitation at room temperature with a low pass filter at $\sim$2.95 eV. }
\end{center}
\end{figure}

From the above discussion, we expect that there is significant radiative recombination in BFO. This is supported by the room temperature time-integrated PL measurement shown in Fig. \ref{f:PL}. The measured PL spectrum consists of a sharp peak at 2.65 eV embedded in a broadband (2.3-2.85 eV) PL signal. The sharp peak agrees well with the reported 2.5-2.8 eV band gap \cite{Kumar2008APL,Palai2008PRB,Xu2009PRB,Ramirez2009PRB} and originates from radiative recombination across the gap. We cannot conclusively identify the origin of the broadband emission from our experiments. However, it is worth noting that the PL spectrum is consistent with the aforementioned "smearing" of the absorption edge to lower energies, with an onset at $\sim$2.2 eV.\cite{Pisarev2009RPB,Xu2009PRB,Ramirez2009PRB}  Although the origin of this edge smearing is still unclear, we note that Ref. \onlinecite{Pisarev2009RPB} attributes it to the presence of low lying electronic structure that starts from $\sim$2.2 eV and originates from lowered crystal symmetry due to lattice distortion, which leads to the enhancement of previously forbidden transitions. Another possibility is luminescence from defect states that are known to exist in BFO films,\cite{Basu2008APL,Hauser2008APL} as is often seen in semiconductors.\cite{Rohit2011Book} However, since edge smearing is observed in most ferrites (in both single crystals and films) and is stronger in materials with larger crystal distortions, it is likely that this effect is an intrinsic property of BFO (and not due to defect states), as suggested in Ref. \onlinecite{Pisarev2009RPB}.  Additional PL measurements performed on BFO films grown on SrTiO$_{3}$ substrates (not shown) support this hypothesis, as broadband emission was also observed, but at much lower energies ($\sim$1.5-2.25 eV)\cite{SheuToPublish}. This implies that the origin of this signal is different than in our single crystal sample, with defect state luminescence a likely possibility. We can thus attribute the second (intermediate) decay time constant in our ultrafast optical data to electrons leaving the conduction band, due both to direct radiative recombination across the gap (corresponding to the sharp peak in the PL spectrum) as well as radiative recombination occurring after carriers relax to another split state lowered in energy by symmetry breaking \cite{Pisarev2009RPB} (corresponding to the broadband emission).

\begin{figure}[tb]
\begin{center}
\includegraphics[width=3.5in]{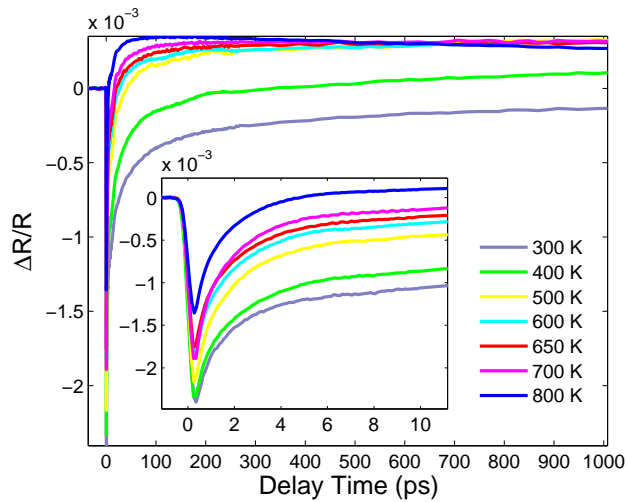}
\caption{ \label{f:SC395nm_T} (color online) Photoinduced transient reflectivity change at various temperatures measured at $\sim$3.18 eV. The inset shows the dynamics on short timescales.}
\end{center}
\end{figure}

Finally, we also performed temperature-dependent experiments to examine the influence of AFM order on carrier dynamics in BFO (Fig. \ref{f:SC395nm_T}). The initial magnitude of the $\Delta$R/R signal and the point in time where the signal changes sign from negative to positive both decrease with increasing temperature. This is consistent with the band gap decreasing in magnitude with temperature (dE$_g$/dT$<$0), since for a fixed pump/probe photon energy, the excess energy of the photoexcited electrons increases with temperature, thus increasing the lattice heating after photoexcited electrons relax to the conduction band minimum through electron-phonon coupling. We estimate the gap change from 300-800 K in our data to be $\sim$0.59 eV,\cite{dEnotes} which agrees with experimental reports and theoretical estimates in Ref. \onlinecite{Palai2008PRB}. The $\Delta$R/R signal at long time delays is thus due to the combined contributions of heat diffusion (positive component of the signal) and carrier recombination (negative component of the signal). For instance, at high temperatures more excess energy is transferred to lattice heating and the late time decay is dominated by heat diffusion, whereas at low temperatures the pump energy is closer to the bottom of the conduction band and the late time decay is dominated by the slow carrier recombination.

We find that the fast time constant is nearly constant over the measured temperature range, while the intermediate time constant becomes faster at higher temperatures. This is consistent with the fact that electron-phonon coupling does not change dramatically with temperature, because the lattice and electronic heat capacities do not significantly change across the Neel temperature. This also further verifies that spin-lattice relaxation does not play a significant role in the observed dynamics,  since we observe no abrupt change in the measured dynamics across $T_{N}$.

In conclusion, we have presented a comprehensive study of carrier dynamics in single crystal BFO from room temperature to across the Neel temperature. We find that the measured $\sim$1 ps fast time constant is due to carrier relaxation via electron-phonon coupling, which has insignificant temperature dependence up to 800 K. We also observe a negative band gap shift at high temperatures, consistent with earlier reports.\cite{Palai2008PRB,Basu2008APL} We attribute the second temperature-dependent relaxation process to photoexcited electrons leaving the conduction band, either through direct radiative recombination or through radiative recombination from other lower lying electronic states, as supported by our PL data. Finally, the slow recovery of the $\Delta$R/R signal at long time delays is due to a combination of heat diffusion and recombination. Our results thus suggest that carrier relaxation in BFO, despite the presence of FE and AFM order, is largely analogous to that in a bulk semiconductor.

This work was performed at the Center for Integrated Nanotechnologies, a U.S. Department of Energy, Office of Basic Energy Sciences user facility and also partially supported by the NNSA's Laboratory Directed Research and Development Program. Los Alamos National Laboratory, an affirmative action equal opportunity employer, is operated by Los Alamos National Security, LLC, for the National Nuclear Security Administration of the U. S. Department of Energy under contract DE-AC52-06NA25396. The work at Rutgers University was supported by DOE grant no. DE-FG02-07ER46382. We thank Diyar Talbayev for his contributions to this work.


\begin{thebibliography}{27}
\bibitem[{dEn()}]{dEnotes}
\bibinfo{note}{We estimate this change using room temperature parameters and
  from the difference between transient reflectivity traces for different
  photon energies near 3 ps.}
\end{thebibliography}
\end{document}